\documentclass[aps, twocolumn,  amsmath, graphicx, latexsym, amsfonts, asmsymb]{revtex4}

\usepackage{epsfig}

\usepackage{epsfig}

\begin{document}

\title{Entanglement swapping of noisy states: A kind of superadditivity in nonclassicality}

\author{Aditi Sen(De)\(^{1,2,3} \), Ujjwal Sen\(^{1,2,3} \),
{\v C}aslav Brukner\(^{4,5} \), 
Vladim\'\i r Bu{\v z}ek\(^{6,7} \), 
and Marek \.Zukowski\(^{1} \)}

\affiliation{\(^{1}\)Instytut Fizyki Teoretycznej i Astrofizyki, Uniwersytet 
Gda\'nski, PL-80-952 Gda\'nsk, Poland}
\affiliation{\(^{2}\)Institut f\"ur Theoretische Physik, Universit\"at Hannover, D-30167 Hannover,
Germany}
\affiliation{\(^3\)ICFO-Institut de Ci\`encies Fot\`oniques, Jordi Girona 29, Edifici Nexus II,
E-08034 Barcelona, Spain}
\affiliation{\(^{4}\)Institut f\"ur Experimentalphysik, Universit\"at Wien,  
Boltzmanngasse 5, A-1090, Austria} 
\affiliation{\(^5\)Institute of Quantum Optics and Quantum Information, Austrian Academy of
Sciences, Boltzmangasse 3, A-1090 Vienna, Austria}
\affiliation{\( ^{6} \)Research Center for Quantum Information, 
Slovak Academy of Sciences,
D\'ubravsk\'a cesta 9, 
845 11 Bratislava, Slovakia} 
\affiliation{\( ^{7} \)Faculty of Informatics, Masaryk University, Botanick\'a 68a,
602 00 Brno, Czech Republic}

\begin{abstract}
We address the question as to whether an entangled state that satisfies local realism will give 
a violation of the same,
after entanglement swapping in a suitable scenario. We consider 
such possibility as a kind of superadditivity in nonclassicality. Importantly, 
it will indicate that 
checking for violation of 
local realism, in the state obtained after entanglement swapping, can be a \emph{method} for 
detecting entanglement in the input state of the swapping procedure.  
We investigate various entanglement swapping schemes, which involve 
mixed initial states.
The strength of violation of local realism 
by the state obtained after entanglement swapping, is compared with the 
one for the input states. 
We obtain  a kind of superadditivity of violation of 
local realism for Werner states,   consequent upon entanglement swapping 
involving Greenberger-Horne-Zeilinger state measurements. 
We also discuss whether entanglement swapping of specific states may be used 
in quantum repeaters with a substantially reduced need to perform the entanglement distillation step. 
\end{abstract}

\maketitle

\def\com#1{{\tt [\hskip.5cm #1 \hskip.5cm ]}}

\newcommand{\tr}{{\rm tr}}

\section{Introduction}
\label{intro}

Quantum nonseparability, in its operational sense, is the existence
of states which cannot be prepared by distant observers acting locally
and without any supplementary quantum channel. So it may seem that
 particles which do not share a common past (i.e., which have
not been acted on by an interaction Hamiltonian) cannot be nonseparable (or entangled). 
Surprisingly however, 
two particles \emph{can} get  entangled even if they do not share a common past. This is 
achieved in entanglement swapping \cite{ZZHE, ZukowskiSwapProc, BVKswap1,
BVKswap2}. The phenomenon
was  experimentally confirmed in \cite{swapexpt}.

\begin{figure}[ht]
\begin{center}
\unitlength=0.5mm
\begin{picture}(150,50)(0,0)
\thicklines
\put(10,15){\line(1,0){55}}
\put(10,15){\circle*{2}}
\put(8,5){Alice}
\put(55,5){Bob}

\thinlines
\put(55,-5){\framebox(40,40)}

\put(56,40) {Measurement}

\thicklines
\put(80,5){Claire}
\put(125,5){Danny}
\put(65,15){\circle*{2}}
\put(130,15){\circle*{2}}
\put(85,15){\line(1,0){45}}
\put(85,15){\circle*{2}}

\end{picture}
\end{center}
\caption{Entanglement swapping between two states.}
\label{twoswap} 
\end{figure}
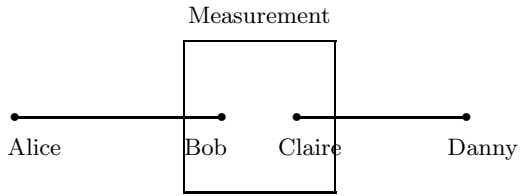
Let us first describe very briefly the phenomenon of entanglement swapping. 
Suppose Alice and Bob share an entangled state. 
Similarly Claire and Danny 
also share some entangled state. See Fig. \ref{twoswap}.
Now the question is as follows: 
Can it be possible that Alice's and Danny's particles become entangled 
without an interaction between their particles?

In Refs. \cite{ZZHE, ZukowskiSwapProc, BVKswap1, BVKswap2}, the authors have shown that the answer is 
Yes. If their partners Bob and Claire (whose particles are entangled with
 the particles of Alice and Danny respectively) come together and make 
a measurement in a suitable basis and communicate their measurement results
clasically (say, by phone call), then Alice's and Danny's particles may become 
entangled.

A simple example of this phenomenon can be seen if  one has
two singlets, 
\(
\frac{1}{\sqrt{2}}(\left|01\right\rangle - \left|10\right\rangle)\), 
  one of which is shared by Alice and Bob, and the other by Claire and Danny. 
(\(\left|0\right\rangle\)
and \(\left|1\right\rangle\) are mutually orthonormal.) Now 
Bob and Claire make jointly
a projection measurement (on 
their parts of the two singlets) in the Bell basis, which is given by 
(with the \(+\) sign applying to states with odd indices)
\begin{eqnarray}
\label{phipm}
\left|B_{1,2} \right\rangle = \frac{1}{\sqrt{2}}(\left|00\right\rangle \pm \left|11 \right\rangle), \nonumber \\
\left|B_{3,4}\right\rangle = \frac{1}{\sqrt{2}}(\left|01\right\rangle \pm \left|10\right\rangle).
\end{eqnarray} 
It is easy to check that if 
 Bob and Claire
 communicate (over a classical channel) the result of their
 measurement to Alice and Danny, they will know 
that they share one of the Bell states given by Eq. (\ref{phipm}).
Note, that depending on the 
measurement results, unitary operations \(\sigma_x\), \(\sigma_y\),  \(\sigma_z\), or \(I\) 
may  be performed by Alice (or Danny) on her (his) qubit 
to obtain just a
singlet. (\(I\) is the identity operator on the qubit Hilbert space, and \(\sigma_x\), \(\sigma_y\),
\(\sigma_z\) are the Pauli matrices.)
 The particles 
of Alice and Danny are completely independent, and nevertheless they share 
entanglement after Bob and Claire's 
 Bell measurement (and sending its outcome to them).
Note that entanglement swapping can be seen as a specific case of teleportation \cite{BBCJPW}. 
In the entanglement swapping process, 
Bob and Claire make a measurement on their systems and send (teleport) the qubit 
(say, Bob's subsystem) 
 through a channel to Danny. 
And after communication to Danny, Alice and Danny share an entangled state. Actually,  
if all the parties agree on the desired output state of the swapping procedure, 
Bob and Claire can communicate their results only to Danny, and Alice does not 
 need to know the content of the communication.

Entanglement in shared multiparty states is a fundamental resource in several quantum communication processes.
However, it is usually a hard problem to detect whether a given state is entangled (see e.g. \cite{bindaas}). 
One way to detect entanglement is to check for violation of local realism. 
However there seems to exist entangled states that does not violate local realism. 
In this paper, we  address the following question:  

\textbf{Question.} \emph{Consider a state that is entangled and yet does not violate local
realism. Is it possible to find some entanglement swapping process, 
after which the swapped state will violate local realism?}
 
We provide a partial answer to this question. To that end, we investigate various entanglement swapping
 schemes which involve mixed entangled states as  initial states.
These will be, for simplicity, modeled as 
partially depolarised states \cite{depolarised}. We will be particularly interested to 
investigate the extent
to which the states resulting out of the entanglement swapping process violate local realism. 
We address the case where the swapping itself (i.e.,
the measurement required for swapping) is perfect.
A study of the complementary situation in which the (multiple) 
swapping is non-perfect, for the {\em realistic} case of the parametric down
conversion process, is given in \cite{ref-amader-pdc}.

The parent states considered here are the ``Werner mixtures'' of certain pure states 
(say \(\left|\psi\right\rangle\), shared between \(n\) partners) and the white noise \cite{depolarised}:
\begin{equation}
\varrho = p\left|\psi\right\rangle\left\langle\psi\right| + (1-p) \varrho_{noise}.
\end{equation}
The parameter \(p\) will
be called here visibility. Clearly it shows to what extent the processes that can be described by
\(\left|\psi\right\rangle\) are operationally visible despite the presence of noise.
It can be associated with the notion of visibility in multipartite interference
experiments.
We shall study the relation of the visibility parameter for the initial states, and the states 
after swapping. This will be done in various configurations:
\begin{enumerate}

\item Chain configuration:
 A  chain of entanglement swappings involving initially a 
sequence of pairs (sharing the parent states) \cite{Dur1,Dur2}.
Bell measurements, i.e. measurements projecting onto the  Bell states given  
by  Eq. (\ref{phipm}),
are performed upon two particles 
 of all adjacent pairs (see Fig. \ref{twoswap} for the case of 
one entanglement swapping with two pairs). This is described in section
\ref{chainswap}.

\item Star configuration: A generalized entanglement swapping involving initially \(N\) parent states 
(each consisting of \(M\) particles).
An \(N\) qubit GHZ state measurement \cite{GHZ} is made on \(N\) qubits, each belonging to a
different state. This is dicussed in section \ref{ringswap}.
GHZ state  measurement projects onto the GHZ basis. 
The \(3\)-qubit GHZ basis, for example, consists of the states
\begin{eqnarray}
\label{GHZbasis}
G_{1,2} = {\frac{1}{\sqrt{2}} (\left | 000\right\rangle \pm \left | 111 \right\rangle )}, \nonumber \\
G_{3, 4} = {\frac{1}{\sqrt{2}} (\left | 100 \right\rangle \pm \left | 011\right\rangle )}, \nonumber  \\
G_{5, 6} = {\frac{1}{\sqrt{2}} (\left | 010 \right\rangle \pm \left | 101\right\rangle )},  \nonumber \\
G_{7,8} = {\frac{1}{\sqrt{2}} (\left | 001 \right\rangle \pm \left | 110\right\rangle )},
\end{eqnarray}
where again the \(+\) sign applies to states with odd indices.
Similarly one may define an \(N\)-qubit GHZ basis by considering the binary decompositions of \(2^N -1\).
  
\end{enumerate}
 
We shall be interested  whether the resulting states after different 
forms of entanglement swapping are nonclassical. 
As our bench-mark of nonclassicality, we shall use the
threshold value of visibility allowing for violation of suitable Bell inequalities \cite{Bell,CHSH}.
That is, we compare the critical visibility 
for violation of local realism of the state 
obtained after entanglement swapping, with the critical visibility for violation in the 
input state (parent state) itself.

We obtain a kind of superadditivity in violation of local realism, 
for the case of Werner states in the Hilbert space of dimension \(2 \otimes 2\), 
consequent to entanglement swapping in a 
specific scenario (section \ref{section_superadditive}).
In the concluding section 
(section \ref{discussion}), we discuss the possibility of the 
existence of such superadditivity for other states, in suitably chosen
configurations of entanglement swapping. There we will come back to the general 
question that we have asked in the beginning. We indicate that checking for violation of local realism in the 
state obtained after entanglement swapping in suitably chosen configurations, can be an efficient 
entanglement witness for the input state. We also discuss the possibility of using entanglement swapping 
with specific states in the so-called ``quantum repeater'' \cite{Dur1,Dur2}, where 
the distillation step may not be required or its requirement may be substantially reduced.

\section{Chain configuration of  entanglement swapping}
\label{chainswap}

In this section, we will  compare the visibilities 
of the input state to that of the swapped state,  in the  
case of entanglement swapping between pairs of states in a chain configuration.
See Fig. \ref{twoswap} for a chain of two pairs.

The chain configuration of entanglement swapping, and the exponential increase of noise in the 
swapped state, has been studied in Refs. \cite{Dur1,Dur2}.  We give this brief discussion here to 
compare with the ring configuration to be considered later.

\subsection{Entanglement swapping between two pairs}

 Consider the \(2 \otimes 2\) dimensional Werner state \cite{Werner1989}
\begin{equation}
\label{Werner}
\rho = p \left|B_1\right\rangle \left\langle B_1 \right| + (1-p) \rho_{noise}^{(2)}.
\end{equation}
(In this paper, we denote 
the completely depolarised state of \(n\) qubits, \(I_n/2^n\),  as \(\rho_{noise}^{(n)}\), where
\(I_n\) is the identity operator of the Hilbert space of \(n\) qubits.) 
The state \(\rho\) is entangled when \(p > {\frac{1} {3}}\), 
but the state violates local realism
only for \(p > {\frac{1}{\sqrt{2}}}\). 
Let Alice (A) and Bob (B) share the state \(\rho\), and let Claire (C) and Danny (D) 
also share such a state \cite{sameparent}. 
 Bob and Claire (who are together) 
make a measurement on their part of the two states, in the Bell 
basis \(\left\{B_{i}\right\}\) given by eqs. (\ref{phipm}) (see Fig. \ref{twoswap}).
We are interested in whether 
the  state of the 
qubits of Alice and Danny after the swap,
violates local realism. 
After the  Bell measurement, if the 
measurement result is \(B_{1}\), the state shared by Alice 
and Danny is  a Werner state of the form
\begin{equation}
 \xi^{(2)}_{AD}
= p^{2} \left|B_1\right\rangle \left\langle B_1 \right| + (1- p^{2})\rho_{noise}^{(2)}.
\end{equation}
Since \(\xi^{(2)}_{AD} \) is  a Werner state, it is entangled for 
\(p > \frac{1}{\sqrt 3}\), but violates local realism when 
\(p > \left(\frac{1}{2}\right)^\frac{1}{4}\). Of course, the same condition is
obtained for the other Bell measurement outcomes.
Therefore the region in which the 
final state \(\xi^{(2)}_{AD}\) violates  Bell inequalities is strictly contained in 
the region in which the initial state \(\rho_{AB}\) has the same property.
We see that there is a region of \(p\), namely  
\(p \in (\frac{1}{\sqrt{2}}, \left(\frac{1}{2}\right)^{\frac{1}{4}})\),
for which the output  state will not be able to show  any
 violation of   local realism (but it is still entangled), whereas 
the input states 
do violate in that region.
Therefore we have a ``loss in the region of violation of local realism'' after  entanglement 
swapping.

\subsection{Chain of \(N\) states: ``Loss'' increases with \(N\)}

This phenomenon of ``loss'' becomes more and more pronounced as the number of swappings 
is increased. Starting with \(N\) initial Werner states in eq. (\ref{Werner}) 
shared between 
\(A_k\) and \(B_k\) \((k = 1, 2, \ldots, N)\),
 the swapped state between \(A_1\)
and \(B_N\) (after Bell measurements performed by 
\(B_1 A_2\), \(B_2 A_3\), \(\ldots\), \(B_{N-1} A_N\))
is again the  Werner state 
\begin{equation}
p^N \left|B_1 \right\rangle
 \left\langle B_1 \right| + 
(1 - p^N)\rho_{noise}^{(2)}.
\end{equation} 
 Hence the swapped state violates local
realism for 
\begin{equation}
p> \left(\frac{1}{2}\right)^\frac{1}{N}.
\end{equation} 
Therefore in the
 case of  
a series of  a large number of entanglement swappings,
the swapped 
state can violate local realism only when initial state is almost pure.

Note that if we consider a chain of  \(N\) Werner states with different visibilities, i.e. if  
\begin{equation}
p_k \left|B_1 \right\rangle \left\langle B_1 \right| 
+ (1 - p_k)\rho_{noise}^{(2)}
\end{equation}
is  shared between 
\(A_k\) and \(B_k\) \((k = 1, 2, \ldots, N)\),
 then the swapped state between \(A_1\)
and \(B_N\) 
is the Werner state 
\begin{equation}
p_1 p_2 \ldots p_N \left|B_1 \right\rangle
 \left\langle B_1 \right| + 
(1 - p_1p_2 \ldots p_N)\rho_{noise}^{(2)}.
\end{equation}
Therefore, again we have that the region of violation of local realism of the swapped state
is strictly smaller than the region of violation of the parent 
states in the \((p_1, p_2, \ldots, p_N)\)-space. 
The former is vanishing
for sufficiently large \(N\).

\section{A star configuration entanglement swapping}
\label{ringswap}

Let us consider entanglement swapping in a different configuration, than that was considered
in section \ref{chainswap}.
Assume a multiparty situation in which initially disjoint subsets of parties share entangled states.
In the next stage,  single representatives of each subset of parties meet together and perform  a 
GHZ-state measurement. The result of the measurement is sent to the remaining parties.
 This procedure results in an entangled state shared by 
them. We shall call this type of entanglement swapping as entanglement swapping in a ``star configuration''.

\subsection{A star swapping between three states} \label{ringthree}

\begin{figure}[ht]
\begin{center}
\unitlength=0.5mm
\begin{picture}(100,80)(0,0)
\thicklines

\put(45,55){\line(-1, 1){20}}
\put(45,55){\circle*{2}}
\put(25,75){\circle*{2}}
\put(21,65){\(B\)}
\put(38,50){\(A\)}

\put(65,55){\line(1, 1){20}}
\put(65,55){\circle*{2}}
\put(85, 75){\circle*{2}}
\put(87,70){\(D\)}
\put(67,50){\(C\)}

\put(55,40){\line(0,-1){30}}
\put(55,40){\circle*{2}}
\put(55,10){\circle*{2}}
\put(58,10){\(F\)}
\put(57, 35){\(E\)}
\thinlines
\put(30,30){\framebox(45,30)}

\end{picture}
\end{center}
\caption{A star configuration swapping.
 A GHZ basis measurement is performed on A, C, and E. This is represented by a box.}
\label{fig_ringthree}
\end{figure}
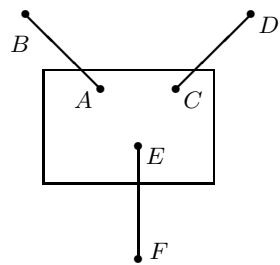
Suppose that pairs AB, CD and EF, each share the same Werner state 
\(\rho = p \left|B_1\right\rangle \left\langle B_1\right|
+ (1 - p) \rho_{noise}^{(2)}\) (see Fig. 2).
A, C,  and E 
come together and
perform a measurement in their \(2\otimes 2 \otimes 2\) dimensional Hilbert space,
in the GHZ basis as given in Eq. (\ref{GHZbasis}).
After the measurement, if \(G_{1}\) clicks, then B, D, F share the state
\begin{eqnarray}
\label{three}
 \xi^{(3)}_{BDF} & = &    
 p^{3}\left | G_{1}\right\rangle \left \langle G_{1}\right| +  (1- p^2) \rho_{noise}^{(3)} \nonumber \\
&  & +   \frac{1}{2} p^2 (1 - p)
(\left | 000\right\rangle \left \langle 000\right| + \left | 111\right\rangle \left \langle 111\right|).
\nonumber \\    
\end{eqnarray}
Other measurement results give the same state upto
local unitary transformations.
This indicates that the amount of violation of local realism, considered 
below for the state \(\xi^{(3)}_{BDF}\), is also attained for the swapped states 
for other measurement results. Thus, although the partners  B, D, F need to know 
the result of the measurement performed by A, C, E, they can keep the output state
in every run of the measurement. 
Note that the swapped 
state now is not a mixture of white noise  and 
\(\left | G_{1}\right\rangle \left \langle G_{1}\right| \) only,
 and this is so whenever \(p \ne 1\).

We  will now use  the Mermin-Klyshko (MK)
inequalities 
to study the violation of local realism 
by the swapped state (see Appendix). 
Let us first 
calculate 
\(\mbox{tr}\left(B_{3}\xi^{(3)}_{BDF}\right)\). (See Eq. (\ref{MK}).)
Suppose that the observables are chosen from the \(x-y\)-plane \cite{why_x-y}.
That is, we choose 
\begin{eqnarray}
\label{x-y}
\sigma_{a_j} = \left|+,\phi_j\right\rangle \left\langle +,\phi_j \right|
                - \left|-,\phi_j\right\rangle \left\langle -,\phi_j \right| \nonumber \\
\sigma_{a{'}_j} = \left|+,\phi{'}_j\right\rangle \left\langle +,\phi{'}_j \right|
                - \left|-,\phi{'}_j \right\rangle \left\langle -,\phi{'}_j \right|,
\end{eqnarray}
where 
\begin{eqnarray}
\left|\pm ,\phi_j\right\rangle = \frac{1}{\sqrt{2}}(\left|0\right\rangle \pm e^{i\phi_j}\left|1\right\rangle) \nonumber \\
\left|\pm ,\phi{'}_j\right\rangle = 
\frac{1}{\sqrt{2}}(\left|0\right\rangle \pm e^{i\phi{'}_j}\left|1\right\rangle). 
\end{eqnarray}

The only term in the state given by (\ref{three}), that will contribute 
to the expression 
\(\mbox{tr}\left(B_{3}\xi^{(3)}_{BDF}\right)\), is the first one, and more precisely, its part given by
\begin{equation}
{\frac{p^3}{2}} (\left|000\right\rangle\left\langle 111\right| + 
\left|111\right\rangle\left\langle 000\right|).
\end{equation}
 (This 
observation would help us in the more general cases that we consider in the succeeding subsections.)

For the  GHZ state \(\left|G_1\right\rangle = \frac{1}{\sqrt{2}}(\left|000\right\rangle 
+ \left|111\right\rangle)\), one has
\begin{equation}
\max \mbox{tr}\left(B_{3}\left|G_1\right\rangle \left\langle G_1\right|  \right)=2,
\end{equation}
and this maximal violation of local realism by the GHZ state
is reached in the \(x-y\)-plane. 
Therefore the maximal value reached by  \(\mbox{tr}\left(B_{3}\xi^{(3)}_{BDF}\right)\),
for any choice of \(\phi_j\) and \(\phi{'}_j\) by the parties, is given by
\begin{equation}
\label{conditionthree}
 \max \mbox{tr}\left(B_{3}\xi^{(3)}_{BDF}\right) = 2 p^{3}.
\end{equation}

Consequently, the state   \(\xi^{(3)}_{BDF}\) violates a MK inequality for 
\(\max \tr(B_3 \xi^{(3)}_{BDF}) >1\),
for
 \begin{equation}
  p > \left(\frac{1}{2}\right)^{\frac{1}{3}} \simeq .7937 .
  \end{equation} 
Our initial Werner state \(\rho\) violates Bell inequalities when 
\begin{equation} 
p > \frac{1}{\sqrt{2}} \simeq .7071.
\end{equation} 
One should compare this with the case of entanglement 
swapping between \emph{two} Werner states, where 
the swapped state gives  violation for 
\begin{equation}
 p> \left(\frac{1}{2}\right)^\frac{1}{4} \simeq .8409.
 \end{equation}

While considering violation of local realism by the 
state \(\xi_{BDF}^{(3)}\), we have used only the Mermin-Klyshko inequalities. 
However, in this case one can also consider  the WWWZB inequalities \cite{WZ, WW, ZB}, 
which are a necessary and
sufficient condition for the violation of local realism by 
the \(N\)-qubit correlations of an arbitrary state
of \(N\) qubits, when there are two settings at each site.

Let us first define the correlation tensor for \(N\)-qubit states. 
An \(N\)-qubit state \(\rho\) can always be written down as
\begin{equation}
\label{eq_general_rho}\frac{1}{2^N} \sum_{x_1, \ldots, x_N = 0, x, y, z} T_{x_1 \ldots x_N}
\sigma_{x_1}^{(1)} \otimes  \ldots \otimes \sigma_{x_N}^{(N)},
\end{equation}
where \(\sigma_0^{(k)}\) is the identity operator and the
 \(\sigma_{x_i}^{(k)}\)'s (\(x_i = x, y, z\)) are the Pauli operators
of the \(k\)-th qubit. The coefficients
\begin{equation}
\label{eq_correlation}
T_{x_1 \ldots x_N} = \tr(\rho\sigma_{x_1}^{(1)} \otimes 
\ldots \otimes \sigma_{x_N}^{(N)}), \quad (x_i = x, y, z)
\end{equation}
are elements of the \(N\)-qubit correlation tensor
\(\hat{T}\) and they fully define the \(N\)-qubit correlation functions of the state \(\rho\).

Consider now the state \(\xi^{(3)}_{BDF}\), obtained via entanglement 
swapping, as given in Eq. 
(\ref{three}). One can 
check that the three-qubit correlation tensor \(\hat{T}\) 
of this state, contains only those terms which are also present for
 the GHZ state \(G_1\). 
Precisely,  the correlation tensor  \(\hat{T}\)  
of \(\xi^{(3)}_{BDF}\), is given by 
\begin{eqnarray}
\hat{T}_{\xi^{(3)}_{BDF}} = && p^3 \hat{T}_{G_1}
\end{eqnarray}
where \(\hat{T}_{G_1}\) is 
the correlation tensor of the GHZ state \(G_1\) 
given by
\begin{eqnarray}
\hat{T}_{G_1} = &&
\vec{x}_1 \otimes \vec{x}_1 \otimes \vec{x}_1 - 
\vec{x}_1 \otimes \vec{x}_2 \otimes \vec{x}_2 \nonumber \\
&& - \vec{x}_2 \otimes \vec{x}_1 \otimes \vec{x}_2
- \vec{x}_2 \otimes \vec{x}_2 \otimes \vec{x}_1, 
\end{eqnarray}
 with \(\vec{x}_1 = \vec{x}\) and \(\vec{x}_2 = \vec{y}\).
Hence, when the quantum correlation function is computed 
by inserting \(\hat{T}_{\xi^{(3)}_{BDF}}\) 
into the generalised Bell inequality
of WWWZB, one gets the value \(2 p^3\). 
This is because for the GHZ state, the value is
\(2\).
This maximal value (\(2p^3\))  is attained for the measurement in the  \(x-y\) plane,
and was  already
 obtained (in Eq. (\ref{conditionthree})) for the state \(\xi^{(3)}_{BDF}\),
 when we considered  the MK inequalities. 
Therefore the state \(\xi^{(3)}_{BDF}\) 
violates local realism for \(p> (1/2)^{1/3}\).      
Moreover, from our considerations of the WWWZB inequalities, we have that 
for lower values of the parameter \(p\), the three-qubit correlations of \(\xi^{(3)}_{BDF}\)
have a local realistic model for two measurement settings at each site.

\subsection{Other forms of the star configuration of swapping}
\label{morering}

In the preceeding subsection, we have shown that  the ``star configuration'' leads to 
stronger resistance to noise admixture than with Bell measurements in the ``chain configuration'' (discussed in 
section \ref{chainswap}).
The parent states that we considered (in the preceeding subsection) were bipartite states. Let us now
consider
the case of entanglement swapping with measurements 
in  a GHZ basis,
when the parent states are multipartite states.

\begin{figure}[ht]
\begin{center}
\unitlength=0.5mm
\begin{picture}(100,80)(0,0)
\thicklines
\put(45,45){\line(-1, 1){30}}
\put(45,45){\circle*{2}}
\put(15,75){\circle*{2}}
\put(15,15){\circle*{2}}
\put(5,70){\(B\)}
\put(5,11){\(C\)}
\put(44,47){\(A\)}
\put(45,45){\line(-1, -1){30}}

\put(65,45){\line(1, 1){30}}
\put(65,45){\line(1, -1){30}}
\put(65,45){\circle*{2}}
\put(95,15){\circle*{2}}
\put(95, 75){\circle*{2}}
\put(97,70){\(E\)}
\put(97,10){\(F\)}
\put(57,47){\(D\)}
\thinlines
\put(30,30){\framebox(51,30)} 

\end{picture}
\end{center}
\caption{A star configuration swapping. 
A Bell measurement is performed on A and D which is denoted by a box.}
\label{tworing}
\end{figure}
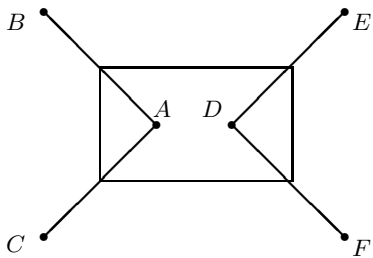

Consider therefore the state
\begin{equation}
\rho_3 =  F \left| G_1 \right\rangle \left\langle G_1 \right| + (1-F)\rho_{noise}^{(3)} 
\end{equation}
where \(\left|G_1\right\rangle = \frac{1}{\sqrt{2}}(\left|000\right\rangle + 
\left|111\right\rangle)\). This state violates local realism for 
\begin{equation}
F > \frac{1}{2}.\end{equation} 
Let two such states be shared between A, B, C and D, E, F,
with A and D placed together (Fig. \ref{tworing}). 
A and D make a Bell measurement on their parts of the two states. After the measurement, 
the resulting state 
 violates the MK inequalities in the \(x-y\)-plane, for 
\begin{equation}
F> \left(2^{\frac{3}{2}}\right)^{-\frac{1}{2}} \simeq .5946.
\end{equation}
Note here that we do not need the explicit form of the state, just like the case for three states. 
The terms in the state that contribute 
to  the violation of MK inequality in the \(x-y\)-plane are 
\(\left |0 \ldots 0\right\rangle \left \langle 1 \ldots 1 \right|  \)
and \(\left |1 \ldots 1\right\rangle \left \langle 0 \ldots 0 \right|  \).

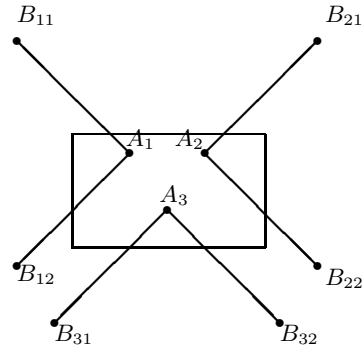
\begin{figure}[ht]
\begin{center}
\unitlength=0.5mm
\begin{picture}(100,100)(0,0)
\thicklines
\put(45,55){\line(-1, 1){30}}
\put(45,55){\circle*{2}}
\put(15,85){\circle*{2}}
\put(15,25){\circle*{2}}
\put(15,90){\(B_{11}\)}
\put(15,21){\(B_{12}\)}
\put(44,57){\(A_1\)}
\put(45,55){\line(-1, -1){30}}

\put(65,55){\line(1, 1){30}}
\put(65,55){\line(1, -1){30}}
\put(65,55){\circle*{2}}
\put(95,25){\circle*{2}}
\put(95, 85){\circle*{2}}
\put(97,90){\(B_{21}\)}
\put(97,20){\(B_{22}\)}
\put(57,57){\(A_2\)}

\put(55,40){\line(-1,-1){30}}
\put(55,40){\line(1,-1){30}}
\put(55,40){\circle*{2}}
\put(25,10){\circle*{2}}
\put(85,10){\circle*{2}}
\put(25,5){\(B_{31}\)}
\put(85,5){\(B_{32}\)}
\put(53, 42){\(A_3\)}

\thinlines
\put(30,30){\framebox(51,30)}

\end{picture}
\end{center}
\caption{A star configuration swapping. \(A_1\), \(B_{11}\), \(B_{12}\) and 
 \(A_2\), \(B_{21}\), \(B_{22}\) and \(A_3\), \(B_{31}\), \(B_{32}\) share noisy GHZ states. All
\(B\)'s are at distant locations but \(A\)'s are in the same lab. 
A GHZ basis measurement is performed by \(A_1\), \(A_2\), and \(A_3\), as depicted in the figure by a box.} 
\label{ringgeneral} 
\end{figure}

With three \(\rho_3\)'s, and a swapping 
in the \(3\)-qubit GHZ basis (given in Eq. (\ref{GHZbasis})) on the 3 qubits (one from each of 
the \(\rho_3\)'s) (see Fig. \ref{ringgeneral}), the MK inequality is
violated in the \(x-y\)-plane for 
\begin{equation}
F> \left(2^{\frac{5}{2}}\right)^{-\frac{1}{3}} \simeq .5612.
\end{equation}

\emph{Thus the following picture is emerging}: Entanglement swapping involving GHZ measurements 
is less fragile (to violation of local realism)
 than  Bell measurements, with repect to the noise admixtures in the initial states.

\subsection{The general  star configuration  entanglement  swapping}
\label{generalring}

We will now generalize the entanglement swapping process in the star configuration.
Consider   the following \(M\)-qubit state:
\begin{equation}
\label{noisyGHZ} \rho_{M} = 
 V \left| GHZ _{M}\right\rangle \left\langle GHZ_{M}\right|
+ (1 - V)\rho_{noise}^{(M)}
\end{equation}
where \(\left |GHZ_{M}\right\rangle = \frac {1}{ \sqrt 2}\left( \left| 0 \right\rangle^{\otimes M}
        + \left| 1\right\rangle^{\otimes M} \right) \).
Take \(N\) copies of  \(\rho_{M}\). 
The \(i\)-th copy \((i=1, 2, \ldots, N)\) is shared between 
\(A_{i}\) and \(B_{i1}, B_{i2}, \ldots, B_{i(M-1)}\). 
We suppose that all \(A_{i}\)s are at the same location of the  observer called Alice. (The schematic 
diagram in Fig. \ref{ringgeneral} is drawn when both  \(N\) and \(M\) are three.)
She makes a measurement in the \(N\)-qubit GHZ basis. 
(See Eq. (\ref{GHZbasis}) for the three qubit GHZ basis.) 
As in the previous cases, we take the observables in the \(x-y\)-plane, i.e., the ones given by  
 Eq. (\ref{x-y}), in the MK inequality. 
Here also we do not need the explicit form of the state. The terms that contribute 
to the violation of MK inequality in the \(x-y\)-plane are 
\(\left |0 \ldots 0\right\rangle \left \langle 1 \ldots 1 \right|  \)
and \(\left |1 \ldots 1\right\rangle \left \langle 0 \ldots 0 \right|  \).
Therefore we
obtain that the resulting \(N(M - 1)\)-qubit state 
violates this inequality for
\begin{equation}
\label{main}
 V > V_{N}^{(M)} \equiv \left(2 ^{\frac{N(M-1)-1}{2}}\right)^{-\frac{1}{N}}.
\end{equation}
This expression is easily obtained once we remember our observation for 
the derivation of Eq. (\ref{conditionthree}) \cite{general_comments}.

\subsection{There is no loss in the asymptotic regime}

\begin{figure}[tbp]
\begin{center}
  \epsfig{figure=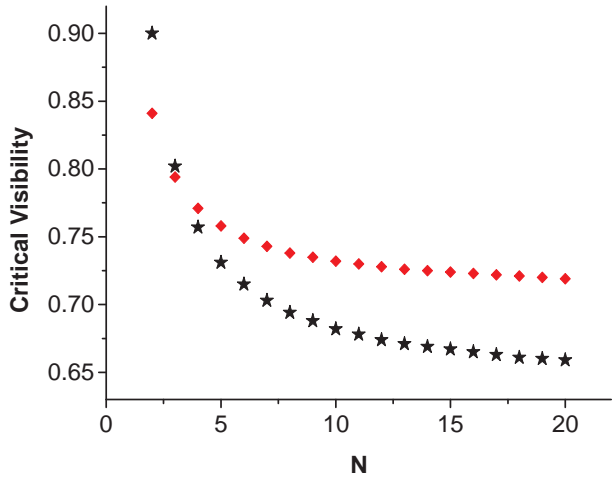,width=0.50\textwidth}
\caption{(a) Plot 
of the critical visibilities \(V_N^{(M)}\) for \(M=2\) (the stars). This is the critical visibility 
required (in the parent state) for violation of local realism by the 
swapped state, in the case when one representative from each of 
\(N\) Werner states (Eq. (\ref{Werner})) come together to perform an entanglement swapping in the \(N\)-qubit 
GHZ basis (see Fig. \ref{fig_comparison}).
 The violation of local realism in the swapped state is considered by using the MK inequalities. 
(b) The critical visibility  \(V_N^f\) (the diamonds) is that for violation of local realism
 by considering functional Bell inequalities in the same \(N\)-qubit swapped state.
The figure shows  the relative monotonic decrease of the two visibilities with  \(N\).}
\label{Plot_of_V_N_for_M_equals_2}
\end{center}
\end{figure}

We remember that our parent state \(\rho_{M}\), as given  in (\ref{noisyGHZ}),
 violates local realism for 
\begin{equation}
 V > \left(\frac{1}{\sqrt{2}}\right)^{M -1}.
 \end{equation}
Note that \(V_{N}^{(M)}\)  (as given by Eq. (\ref{main}))
is monotonically decreasing with respect to 
\( N \).
A plot of the critical visibility \(V_N^{(M)}\) 
for \(M=2\), that is  the visibility obtained when the swapping in a star configuration is performed on
\(N\) number of copies of two qubit Werner states, is given in Fig. \ref{Plot_of_V_N_for_M_equals_2}.
It clearly shows the monotonic decrease  of the critical visibility in \(N\).

Thus the system is surprisingly robust to noise admixture, with respect to violation of local realism in the 
following sense. The amount of (white) noise that the parent state can afford so that 
the state after entanglement swapping still
violates local realism, increases monotonically as we consider swapping between higher 
number of parties, in a star configuration.
Moreover, one has
\begin{equation}
V_{N}^{(M)} \rightarrow \left(\frac{1}{\sqrt{2}}\right)^{M -1} 
\mbox{ as }  N \rightarrow \infty  .
\end{equation}
 This shows 
 that the amount of noise that the parent state can afford, so that the state obtained after entanglement swapping 
violates the MK inequality, 
in the asymptotic limit of 
arbitrarily large number of subsystems (in the way considered above, that is in the star situation),
\emph{coincides} with the amount of noise that can be afforded  by  the parent state itself to violate local realism.
The loss in the
region of violation of local realism is more and more recovered as we consider entanglement 
swapping between higher and higher number of parties
and ultimately in the asymptotic limit, there is \emph{no} loss in the region of 
violation of local realism.

In this general situation,  the state obtained after performing the 
entanglement swapping, is an incoherent mixture
of some product states and a ``weakened'' GHZ state (i.e., a GHZ state admixed with white noise). 
The product states contribute only to the \(T_{z \ldots z}\)  component of the 
the correlation tensor  \(\hat{T}\)  (cf. Eq. (\ref{eq_correlation})) of the state obtained after entanglement 
swapping. Here we have considered violation of the MK inequalities (by the state 
obtained after swapping) only in the \(x-y\)-plane. This is because the gradual reduction 
of loss of the region of violation of local realism after entanglement swapping,
and disappearance
of this loss asymptotically, is already obtained in this plane. However we do not rule 
out a faster reduction of loss if all the WWWZB inequalities are considered.

From the 
perspective of 
the recent works 
indicating that Bell inequality violation is a signature 
of ``useful entanglement" \cite{ScaraniGisincryptoprl, AcinBelldistill, MZCOMPLEXITY, amader_crypto}, 
our result here can 
be also viewed as showing (in a particular case) that in an entanglement 
swapping process, this useful entanglement is lost, but this loss may be asymptotically vanishing.
Below in sections \ref{section_ring_functional} and \ref{section_superadditive}, 
we will show that useful entanglement can even be ``gained'' in an entanglement 
swapping process, and this gain can be possible even without going into the asymptotic regime.
Here by ``gain'', we mean a situation in which the swapped state violates local realism, 
even when the parent 
state does not violate. That is, there exists values of the visibility \(V\), for which 
the parent states do not, while  the swapped state does violate
Bell inequalities
after performing the swap.
We will perform the swapping in a star configuration.

\subsection{Star entanglement swapping in the light of functional Bell inequality
}
\label{section_ring_functional}

The Bell inequalities we have considered upto now are the ones 
in which there is only a finite number of
(in fact, two) settings
per local site. However there are Bell inequalities in which one may consider even a 
continuous range of settings of the local apparatus, as described in 
Appendix \ref{functional_Bell_Zukowski}.

Let us consider violation of local realism by the swapped state  as revealed by a 
 functional Bell inequality.
For simplicity, let us consider the parent states to be a two-qubit state, although all our 
considerations can be generalised to a parent state of higher number of qubits.
Suppose therefore that the Werner state, given by Eq. (\ref{Werner}), is shared between 
two parties, A and B. Numerical calculations have indicated that the Werner state violates local 
realism
for \(p> \frac{1}{\sqrt{2}}\) even for a high number of settings  per observer
\cite{Kasznumerical,maya2}. 
 We use as a working hypothesis that \(p= \frac{1}{\sqrt{2}}\) is indeed the threshold value below 
which there exist an explicit local realistic model which returns the quantum 
predictions for the continuous range of settings.

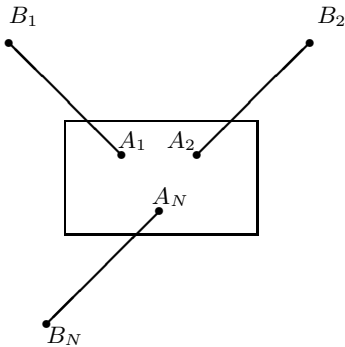
\begin{figure}[ht]

\begin{center}
\unitlength=0.5mm
\begin{picture}(100,100)(0,0)
\thicklines
\put(45,55){\line(-1, 1){30}}
\put(45,55){\circle*{2}}
\put(15,85){\circle*{2}}

\put(15,90){\(B_{1}\)}

\put(44,57){\(A_1\)}

\put(65,55){\line(1, 1){30}}

\put(65,55){\circle*{2}}

\put(95, 85){\circle*{2}}
\put(97,90){\(B_{2}\)}

\put(57,57){\(A_2\)}

\put(55,40){\line(-1,-1){30}}

\put(55,40){\circle*{2}}
\put(25,10){\circle*{2}}

\put(25,5){\(B_{N}\)}

\put(53, 42){\(A_N\)}

\thinlines
\put(30,34){\framebox(51,30)}

\end{picture}
\end{center}

\caption{
\(N\) 
Werner states are distributed between \(A_i\) and \(B_i\) (for \(i = 1, 2,\)
\(\ldots, N\)) and a GHZ basis measurement is performed 
at \(A_1, A_2, \ldots, A_{N}\).}
\label{fig_comparison}
\end{figure}

Consider now the ``star'' configuration 
 described before, where \(A_1 B_1\), \(A_2 B_2\), \(\ldots\),  \(A_N B_N\)
share \(N\) Werner states, each given by Eq. (\ref{Werner}) (see Fig. \ref{fig_comparison}). 
The \(A_i\)'s come together and perform a 
measurement in the \(N\)-qubit GHZ basis (see Eq. (\ref{GHZbasis})) as has been discussed previously. 
We will now consider violation of local realism of the swapped state, by using the 
functional Bell inequalities.

Consider the local observable at the \(j\)th 
location to be 
 \begin{equation}
 \label{observable}
  \sigma_{a_j}(\phi_{j}) = \left| +, \phi_j \right\rangle \left\langle +, \phi_j \right|  -
 \left| -, \phi_j \right\rangle \left\langle -, \phi_j \right|,
 \end{equation}
 where
 \begin{equation}
 \label{eigvec}
 \left|\pm,\phi_j\right\rangle =
 \frac{1}{\sqrt{2}}\left(\left|0\right\rangle \pm
 e^{i\phi_{j}}\left|1\right\rangle\right).
 \end{equation}
  The aggregate \(\xi_j\) of local parameters, in the functional Bell inequality, at the \(j\)th site
is just the single parameter \(\phi_j\) here.
One can now easily show that the swapped state 
violates local realism (by violating the functional Bell inequality with the observables
as defined in Eq. (\ref{observable}))
for \cite{MZfunctional,maya100,MZ100}
\begin{equation}
\label{mainf}
p > V^f_N \equiv  \frac{2}{\pi}\left(2\right)^{\frac{1}{N}}.
\end{equation}
\(V^f_N\) is plotted in Fig. \ref{Plot_of_V_N_for_M_equals_2} 
and compared with the critical visibility \(V_N^{(2)}\),
for the same swapped state, but when violation of the MK inequalities are considered.

\subsection{A kind of superadditivity  for Werner states}
\label{section_superadditive}

There is an important consequence of the relation (\ref{mainf}). For \(N\geq 7\), the critical visibility
\(V^f_N\) is strictly less  than \(\frac{1}{\sqrt{2}}\), which is the critical visibility
 for the Werner state to violate local realism 
(on the basis of the multi-settings numerical results
in Ref. \cite{Kasznumerical}).

Therefore, in the process of entanglement swapping, there seems to be a kind of superadditivity 
 with respect to violation of local realism. 
Suppose Alice shares \(7\) Werner states with \(7\) Bobs, \(B_1\), \(\ldots\), \(B_7\). 
Each of the states is 
given by Eq. (\ref{Werner}),  with a visibility \(p \in (\frac{2}{\pi},\frac{1}{\sqrt{2}})\). 
Such Werner states do not violate local realism
\cite{Kasznumerical}.
 Now suppose that Alice 
makes a measurement in the generalised GHZ basis
 and communicates her result to all the Bobs, over a 
classical channel. The state created at the Bobs, violates 
local realism (by violating 
the functional Bell inequality, as discussed above) for (cf. (\ref{mainf}))
\begin{equation}
\label{qwerty}
p>\frac{2}{\pi}(2)^{\frac{1}{7}} \simeq .7029, 
\end{equation}
which is strictly less than \(\frac{1}{\sqrt{2}} \simeq .7071\). 
Yet for such visibilities which are lower than
\(\frac{1}{\sqrt{2}}\), any \emph{single} pair of particles shared between Alice and any one Bob,
will not be able to violate local realism.
(Recall that  we 
have assumed that taking more settings at each site does not help to 
improve the critical visibility 
of violation of local realism by the Werner state \cite{Kasznumerical}.) 
It is in this sense that we obtain  a kind of ``superadditivity''
in  violation of local realism.

For sufficiently large \(N\), 
\begin{equation}
V^f_N \rightarrow \frac{2}{\pi} \simeq .6366.
\end{equation}

\emph{Let us note here a surprising coincidence.}   An explicit construction of
 local hidden variable model 
for the Werner state exists (till date) for all possible  projection measurements in a plane
by the two parties, for just \(p\leq \frac{2}{\pi}\) \cite{LarssonPLA,maya101}.

It must be stressed  that the kind of superadditivity obtained here is not related to
a distillation protocol \cite{huge}. 
As distinct from a distillation protocol, we do not consider measurements depending 
on previous measurements. Also in our case,  the Alices are together while the Bobs can be 
far apart. Collective operations are required on both ends in the usual distillation protocols. 
Both the recurrence method and the one-way hashing method \cite{huge, purificationIBM} require CNOT operations 
at both ends, which is not possible in our case, as the Bobs are not together. 
The distillation protocol (for all entangled states of 
two qubits) in Ref. \cite{Horodecki_puri} starts with a filtering operation, but must 
be subsequently followed by the recurrence method, which is again not allowed in our case. 
The situation is similar for the protocol in Ref. \cite{Deutsch}. In the distillation protocol 
presented in Ref. \cite{Maneva_Smolin}, measurements on more than two copies of the input are required. 
The experimentally feasible protocol given in Refs. \cite{vienna,vienna2} 
also requires collective operations at both ends.
There is a further difference 
of the entanglement 
swapping protocol considered in this paper, with entanglement distillation protocols \cite{huge}. 
As we have noted before (just after Eq. (\ref{three})), in our entanglement swapping scheme,
the receivers of the swapped state (the \(B_i\)'s in Fig. \ref{fig_comparison}) need to 
know the result of the measurement performed by the \(A_i\)'s. However, in 
contrast to the distillation protocols, they do not need to discard the swapped state for 
some measurement results. In a distillation protocol, discarding some of the outputs is absolutely 
essential, as  entanglement cannot increase under local actions.

In Ref. \cite{Peressuperadditivity}, two Werner states shared by  \(A_1B_1\) and \(A_2B_2\), respectively,
 are  shown to violate local realism, although the individual states are non-violating.
 But in Ref. \cite{Peressuperadditivity},  collective 
tests are required at both ends. That is, both \(A_1\) and \(A_2\), and \(B_1\) and \(B_2\) 
are required to be together. In our case, although the Alices must be together, 
the Bobs are   separated.
Therefore the  ``superadditivity'' reached in this subsection 
is of a different kind than the one  in Ref.
\cite{Peressuperadditivity}.

\section{Entanglement swapping in quantum repeaters}
\label{Gisin state}

In the preceeding section,  we have obtained an example in which
 the initial state has a  local realistic model, but surprisingly, after   
entanglement swapping, the resulting state can violate local realism. This was obtained 
with the initial states as Werner states. However these results were obtained with the star configuration, for which the 
entanglement swapping process, swaps multifold two particle entanglement into a multiparticle entanglement. 
That is the initial and the final enatangled states apply to a different number of qubits.

Therefore we shall now ask a different question. Consider two pairs of qubits, each pair independent of 
the other, and both in an identical quantum state  $\rho^{in}$. That is,  say Alice and Bob share $\rho^{in}_{AB}$,
 which is formally identical with the state shared by Claire and Danny $\rho^{in}_{CD}$. 
Is is possible that the entanglement swapping process, involving a (two-qubit) )Bell measurement jointly by Bob 
and Claire may lead to a new state, $\rho^{out}_{AD}$ shared by Alice and Danny, which has the property that it violates 
local realism more strongly than each of the initial states?
 Note, that now we start with two (identical) two-qubit states, and end with another two-qubit state. 
 The properties of entanglement of the initial and the final state can now be compared directly.

Therefore,  we consider below the case of two qubit entaglement in the initial states, two-qubit Bell-state measurements, 
and two qubit final states. As we shall see one can find specific initial states which after entanglement swapping leads to 
two-qubit state which violates local realism more 
robustly than the initial ones.

Consider the  initial 
state 
\begin{equation}
\label{state_gamma}
\ \rho_{\lambda} = \lambda \left| \psi \right\rangle \left\langle \psi \right| + 
{\frac {1-\lambda} {2}}(\left|00\right\rangle \left\langle 00\right| + \left|11\right\rangle \left\langle 11\right|),
\end{equation}
where \(\left|\psi\right\rangle = a \left|01\right\rangle - b \left|10\right\rangle\)
(and \(\lambda > \frac{1}{2(1-ab)}\)). It  is entangled whenever \(\lambda > 1/(1+2ab)\) \cite{Peres-Horodecki}. For 
\(\lambda \leq 1/(1+a^2b^2)\), this state does not violate any Bell 
inequality. 
However, despite the fact that for \(\lambda \in  (\frac{1}{(1+ 2ab)},\frac{1}{(1+a^2b^2)})\), the state
\(\rho_\lambda\) can be modelled with local hidden variable models, it was shown in Ref. \cite{Gisin1996-filters} 
that after a suitable local filtering operation \cite{gop-churi}, the resulting state violates local realism.

Thus, 
Alice and Bob share a state \(\rho_\lambda\), and so do the other two.
After a Bell measurement  performed by Bob and Claire, if the measurement result is \(\left|B_{1}\right\rangle\),
the final state  of AD
is  
\begin{eqnarray}
\label{eq_exciting}
\xi^{(\rho_{\lambda})}_{AD} & = & 
{\frac{1}{A}}[{\frac{\lambda^2a^2b^2}{2}} \left|B_1 \right\rangle \left\langle B_1 \right| \nonumber \\
&+& {\frac{(1-\lambda)^2}{8}})(\left|00\right\rangle\left\langle 00\right| + 
\left|11\right\rangle\left\langle 11\right|) \nonumber \\
              & + &{\frac{\lambda(1 - \lambda)}{2}} (a^2\left|01\right\rangle\left\langle 01\right| + 
b^2\left|10\right\rangle\left\langle 10\right|)],
\end{eqnarray}
where \(A = 
\lambda^2a^2b^2 + (1-\lambda^2)/4\). This state violates local realism for 
\begin{equation}
\lambda > \frac{1}{\sqrt{1+4(\sqrt{2}-1)a^2b^2}}.
\end{equation}
Therefore the  region of 
violation of local realism for the swapped state \(\xi^{(\rho_{\lambda})}_{AD}\) is not greater 
than that for the parent states (that is, for \(\rho_\lambda\)).
Despite this fact, the \emph{amount} of violation of local realism is greater in the state 
\(\xi^{(\rho_{\lambda})}_{AD}\), as compared to that in the initial state 
\(\rho_\lambda\)), for some ranges of the parameters.

We will now indicate that it is potentially possible to 
use the state \(\rho_\lambda\) (of Eq. 
(\ref{state_gamma})) in a quantum repeater, where there may be a reduced need to perform the 
entanglement distillation step. The entanglement of formation \cite{huge, Wootters_kapano}
of the state \(\rho_\lambda\) is given by 
\begin{eqnarray}
E_{\rho_{\lambda}} = H\left( \frac{1 + \sqrt{1 - C_{\rho_\lambda}^2}}{2} \right),\\
C_{\rho_\lambda} = \max \{0, (1 + 2 a b) \lambda -1 \},
\end{eqnarray}
where \(H\) is the binary entropy function given by 
\(H(x) = - x \log_2 x - (1 - x) \log_2 (1-x)\).
The entanglement of formation of the state \(\xi^{(\rho_\lambda)}\) (given in Eq. (\ref{eq_exciting})), 
obtained after entanglement 
swapping with two \(\rho_\lambda\) can also be similarly calculated by using the 
prescription given in Ref. \cite{Wootters_kapano}.  
One can then check that there exist ranges of the parameters \(a\) and \(\lambda\), 
for which the entanglement of formation of the state \(\xi^{(\rho_{\lambda})}\)
is greater than that in the state \(\rho_\lambda\). 
It may be possible to use this phenomenon in a  
quantum repeater, in which the need to perfom the distillation step is 
substantially reduced. We will follow this up in a later publication. 

Note that the ``superadditivity'' reported
in section 
\ref{section_superadditive} (for the Werner states),
is of a different kind than in Ref. \cite{Gisin1996-filters}.
Importantly, note here that the superadditivity reported in section \ref{section_superadditive}, is 
 for Werner states. 
And at least for a single copy of a Werner state, one cannot reproduce the kind of 
``self-superadditivity'' by using  local filtering operations  \cite{FrankMichael}, as was done 
in \cite{Gisin1996-filters}. Moreover, for the case of Werner states (section \ref{section_superadditive}),
although we do require postselection (just as in Ref. 
\cite{Gisin1996-filters}), all the postselected cases lead to the same result of increased nonclassicality
(in contrast to that in Ref. \cite{Gisin1996-filters}).

\section{Discussion}
\label{discussion}

To conclude, we have shown an example of entanglement swapping process, 
in which although the initial state has a  local realistic model,  after performing  
entanglement swapping, the final swapped state can violate local realism. This was obtained 
by using the initial states as Werner states. We regard this as a kind of superadditivity 
for Werner states. We have also considered another family of states, in which we have shown 
that the amount of violation of local realism,
as also amount of entanglement (as quantified by entanglement of formation) is increased after entanglement swapping.

Coming back to the general question posed in the Introduction,
it may be true that it is a generic feature that 
an entangled state which satisfies local realism,
 will violate local realism after a suitable entanglement swapping procedure. 
If this is true, then this method can be used to detect entanglement in the laboratory. Suppose 
Alice and Bob who are in a different locations, share some state. They want to find out whether their shared state 
is entangled or not. One way is to perform a Bell experiment and find whether their state 
violates local realism. If the state violates local 
realism, then they conclude that their state is entangled. If the state does not violate local realism, they cannot
infer anything about the entanglement of the state. However Alice and Bob can 
 apply the method discussed in this paper.
They can perform entanglement swapping on some copies of the state in a suitable configuration, and then check 
whether the resulting state violates local realism. If yes, then they can infer that the input state was 
entangled. It is interesting to find out 
the most general class of states whose entanglement can be detected in this way.

 As we have noted earlier, in general,  our schemes of 
entanglement swapping in different configurations are not 
``distillation'' \cite{huge}. Take for example the ``star'' configutation considered in section
\ref{ringswap} (see Fig. \ref{fig_comparison}).
There,  the
 parties \(B_1\), ..., \(B_N\), 
do not share any
 entanglement before swapping. So, they simply do not have entanglement
 before, and thus cannot ``distill'' it. Our scheme is entanglement
 distribution rather than entanglement distillation. However there is a
 way to see our scheme also as a distillation one.

 For example, in the case of a chain of two pairs of entangled particles
 (A-B and C-D), if Alice has particle A and Bob has particles B, C and D,
 one can consider our scheme as entanglement distillation. Then Bob
 performs Bell-type analysis on particles B and C and projects particles A and
 D on a new entangled state. It is a distillation because Alice and Bob
 had previously shared  entanglement in A-B and after swapping has entanglement in A-D (see 
\cite{BVKswap1,BVKswap2}).  
To see this as a distillation scheme, we must see whether the output in A-D is more entangled 
than the input in  A-B. 
If that is true in some cases, then only a subensemble of swapped pairs
 will be more entangled than the parent pairs.
Others must be
less entangled, as entanglement cannot increase under local operations. 
Interestingly, there exist states for
which an increase of entanglement is possible after entanglement swapping (as shown in Sec. \ref{Gisin state}). This means that for those
states, a ``quantum repeater'' \cite{Dur1,Dur2} may potentially be based only on entanglement swapping. Note
that for the Werner states, one needs both  entanglement swapping \emph{and} entanglement distillation 
(see \cite{Dur1,Dur2}). 
In Sec. \ref{Gisin state}, we indicated a possible candidate for such a phenomenon.

Other entanglement swapping schemes considered in this paper, 
such as the star configuration, can also be considered as
 a ``distillation'' scheme in the following sense. Consider Fig. \ref{fig_comparison} with \(N=3\) (for example),
 and imagine that party \(B_1\) has particle \(B_1\), party \(B_2\) has  particle \(B_2\), and
 party \(B_3\) has particles  \(A_1\), \(A_2\), \(A_3\) and \(B_3\). 
Then, indeed the three parties
 shared two-particle entanglement even before swapping, and the swapped state 
contains genuine three-particle entanglement \cite{genuine1,genuine2}. 
So our scheme
 might not only be a kind of distillation, but also a procedure which can
 transform one type of entanglement  to
 another one (two-particle to three-particle entanglement, in our example).

Finally, it is intriguing to find out whether there exists a Bell inequality for which 
the superadditivity of Werner states considered in this paper, 
can be explained in the following way. Let us consider the 
case of superadditivity for \(7\) Werner states (section \ref{section_superadditive}).
 In the star configuration, there are 
therefore 8 partners who share these states (see Fig. \ref{fig_comparison}). It will be interesting if there
exists a 
Bell inequality
for 8 partners such that the 7 Werner states shared by them will violate the inequality 
for \(p> \frac{2}{\pi}(2)^{\frac{1}{7}}\) (see Eq. (\ref{qwerty})).

\begin{acknowledgments}

A.S. and U.S. acknowledge support from the University of Gda{\'n}sk, Grant
Nos. BW/5400-5-0256-3 and BW/5400-5-0260-4, the Alexander von Humboldt
Foundation, the DFG (SFB 407, SPP 1078, 432 POL), the EC Program QUPRODIS,
the ESF Program QUDEDIS, and EU IP SCALA.
C.B. was supported by the Austrian Science Foundation (FWF) Project No.
SFB 1506 and by the European Commission (RAMBOQ).
VB acknowledges the
support from the EU projects QUPRODIS and QGATES.
MZ is supported by 
Professorial Subsidy of Foundation for Polish Science
and by MNiI
Grant No. PBZ-MIN-008/ P03/ 2003. 
\end{acknowledgments}

\appendix
\section{Bell inequalities}
\label{tools}

For obtaining  violation of local realism 
by the swapped state, we have considered two different types of multipartite Bell inequalities:
multiparticle Mermin-Klyshko inequalities
\cite{Merminineq, Ardehali, Klyshko, RoySingh, Bechman}
(subsection \ref{sectionMK}),
and the functional Bell inequality \cite{MZfunctional} (subsection \ref{functional_Bell_Zukowski}).

\subsection{The Mermin-Klyshko inequalities}
\label{sectionMK}

A Bell operator for the so-called Mermin-Klyshko (MK) inequality for \(N\) qubits (shared between observers 
\(A_1\), \(A_2\), \(\ldots\), \(A_N\))
can be defined recursively as \cite{GisinScarani}
\begin{equation}
\label{MK}
 B_{k} = \frac{1}{2} B_{k-1} \otimes ( \sigma _{a_k} + \sigma _{a{'}_k}) 
 + \frac{1}{2} B{'}_{k-1} \otimes ( \sigma _{a_k} - \sigma _{a{'}_k}), 
\end{equation}
with \(B{'}_{k}\) obtained from \( B_{k}\) by interchanging \(a_{k}\) and \(a_k{'}\),
and 
\begin{equation}
B_1 = \sigma_{a_1} \quad \mbox{and} \quad B{'}_{1} = \sigma_{a{'}_{1}}.
\end{equation}
The party \(A_j\) is allowed to choose between 
the measurements \(\sigma_{a_j}\) and \(\sigma_{a{'}_j}\). 
Here \(\vec{a_j}\) and \(\vec{a{'}_{j}}\)
are two three-dimensional unit vectors (\(j = 1, 2, \ldots, N\)), and for example,
 \(\sigma_{a_j} = \vec{\sigma}.\vec{a_j}\), \(\vec{\sigma} = (\sigma_x, \sigma_y, \sigma_z)\).

An \(N\)-qubit state \(\eta\)  
violates
 MK inequality if 
\begin{equation}
\mbox{tr}\left(B_{N}\eta\right) > 1. 
\end{equation}

\subsection{The functional Bell inequalities}
\label{functional_Bell_Zukowski}

To study the violation of local realism of the swapped state, we will (along with 
the MK inequalities) also consider the functional Bell inequalities \cite{MZfunctional}.

The functional Bell inequalities \cite{MZfunctional}
 essentially follow from
 a simple geometric observation that
 in any real vector space, if for two vectors \(h\) and \(q\) one has
 \(\left\langle h \mid q\right\rangle < \parallel q \parallel^2\),
 then this
 immediately implies that \(h \ne q\). In simple words, if the
 scalar product of two vectors has a lower value than the length
of one of them, then the two vectors cannot be equal.

 Let \(\varrho_N\) be a state shared between \(N\) separated parties.
 Let \(O_n\) be an arbitrary observable measured at the \(n\)th location (\(n=1,\ldots,
 N\)).
 The quantum mechanical prediction \(E_{QM}\) for the
 correlation in the state \(\varrho_N\), when
 these observables are measured, is
 \begin{equation}
 \label{EQM}
 E_{QM}\left(\xi_1, \ldots, \xi_N\right) = \tr \left(O_1 \ldots O_N \varrho_N\right),
 \end{equation}
 where \(\xi_n\) is the aggregate of the local parameters at the \(n\)th site.
 Our objective is to see whether this prediction can be reproduced
 in a local hidden variable theory. A local hidden variable correlation
 in the present scenario must be of the form
 \begin{equation}
 \label{EHV}
 E_{LHV}\left(\xi_1, \ldots, \xi_N\right) = \int d\lambda \rho (\lambda) \Pi_{n=1}
 ^{N}
 I_{n} ( \xi_{n}, \lambda),
 \end{equation}
 where \(\rho(\lambda)\) is the distribution of the local hidden variables and
 \( I_{n}(\xi_{n}, \lambda) \) is the predetermined measurement-result of the
 observable
 \(O_n(\xi_n)\) corresponding to the hidden variable \(\lambda\).

 Consider now the scalar product
 \begin{eqnarray}
 \left\langle E_{QM}\mid E_{LHV}\right\rangle =  \nonumber \\
 \int  d\xi_1 \ldots
 d\xi_N 
  E_{QM}\left(\xi_1, \ldots, \xi_N\right)
 E_{LHV}\left(\xi_1, \ldots, \xi_N\right),
\label{EQMEHV} 
\end{eqnarray}
 and the norm
 \begin{equation}
 \label{normEQM}
 \parallel E_{QM} \parallel^2 =
 \int  d\xi_1 \ldots d\xi_N
 \left(E_{QM}\left(\xi_1, \ldots, \xi_N\right)\right)^{2}.
 \end{equation}
 If we can prove that a strict inequality holds, namely
 for all possible \(E_{LHV}\), one has 
\begin{equation}
\label{star}
\left\langle E_{QM}\mid E_{LHV} \right\rangle \leq B,
\end{equation}
with the number \(B < \parallel E_{QM} \parallel^2\), we will immediately have
 \(E_{QM} \ne E_{LHV}\), indicating that the correlations in
 the state \(\varrho_N\) are of a different character than in
 any local realistic theory. We then could say that the state \(\varrho_N\) violates
 the ``functional" Bell inequality (\ref{star}), as this 
Bell inequality is expressed in terms
 of a typical scalar product for square integrable functions. Note that the 
value of the product depends on a
 continuous
range of parameters (of the measuring apparatuses) at each site.


\begin{thebibliography}{10}
\bibitem{depolarised}A partally depolarised
state is a mixture of the state with white noise, the normalized identity operator of the
Hilbert space corresponding to the state.
\bibitem{why_x-y}We choose the \(x-y\)-plane, as
 one gets maximal violation of local realism, for GHZ states, in this plane.
\bibitem{maya2}In Ref. \cite{Kasznumerical}, it has been 
numerically shown that the Werner state starts 
violating
local realism from \(p = \frac{1}{\sqrt 2}\) (the threshold 
obtained for \(2\) settings per site), even with upto \(8\) settings per site and with 
some assumptions,  
also upto \(14\) settings per site. That is for lower values of \(p\), there exists a local hidden variable 
model for these number of settings.
\bibitem{maya100}The term contributing to  \(\parallel E_{QM} \parallel^2 \) and 
\(\left \langle E_{QM}|E_{LHV}\right\rangle\) is the same term that contributed in the calculation 
for MK inequality. 
One obtains \(\parallel E_{QM} \parallel^2 = \frac{p^N}{2} (2 \pi)^{N}\) and
\( \left \langle E_{QM}|E_{LHV}\right\rangle \leq 4^N\) (see eqns. (20-23) of
 Ref. \cite{MZfunctional} or eqns. (A9-A18) of Ref. \cite{MZ100}). Therefore the state violates 
the functional Bell inequality whenever \[ \left \langle E_{QM}|E_{LHV}\right\rangle \leq 4^N
< \parallel E_{QM} \parallel^2 = \frac{p^N}{2} (2 \pi)^{N},\]  that is whenever 
\(p> V_N^f \equiv \frac{2}{\pi} 2^{\frac{1}{N}}\).
\bibitem{gop-churi} Since the required operations are local, one may interpret this result as a sort of 
``self-superadditivity'',
because one has  here the process: classical state + local filtering = nonclassical state.
\end{thebibliography}
\end{document}